\documentclass[12pt]{article}
\usepackage{color,amsmath,amssymb,amsthm,bm,epsf,epsfig,enumerate,alltt,latexsym,graphicx}
\usepackage[toc,page,titletoc]{appendix}
\usepackage[T1]{fontenc}
\usepackage{mathabx,textcomp,stmaryrd}
\usepackage{comment}
\newcommand{\E}{\ensuremath{\mathrm{E}}}
\newcommand{\var}{\ensuremath{\mathrm{Var}}}
\newcommand{\pr}{\ensuremath{^{\prime}}}
\newcommand{\id}{\ensuremath{\mathrm{I}}}
\renewcommand{\P}{\ensuremath{\mathbf{P}\!}}
\newcommand{\N}{\ensuremath{\mathbf{N}}}
\renewcommand{\sp}{\ensuremath{\mathrm{sp}}}
\newcommand{\tr}{\ensuremath{\mathrm{tr}}}

\newcommand{\mbk}{\ensuremath{\mathbb{K}}}
\newcommand{\1}{\ensuremath{\bm{1}}}
\newtheorem{propn}{Proposition}
\newcommand{\bc}[1]{\ensuremath{\mathbf{\mathcal{#1}}}}
\newcommand{\imp}{\ensuremath{\Longrightarrow}}
\newcommand{\beqn}{\begin{equation}}
\newcommand{\eeqn}{\end{equation}}
\newcommand{\beqns}{\begin{equation*}}
\newcommand{\eeqns}{\end{equation*}}
\newcommand{\beqna}{\begin{eqnarray}}
\newcommand{\eeqna}{\end{eqnarray}}
\newcommand{\beqnas}{\begin{eqnarray*}}
\newcommand{\eeqnas}{\end{eqnarray*}}
\newcommand{\nnlf}{\nonumber\\}

\begin{document}
\begin{center}
\begin{Large}
{\bfseries Three Properties of $F$-Statistics}\\{\bfseries for Multiple Regression and ANOVA}\end{Large}\\
\vspace{.5cm}by Lynn R. LaMotte\footnote{School of Public Health, LSU Health, New Orleans, LA,  {\tt llamot@lsuhsc.edu}}
\end{center}

\begin{abstract}
This paper establishes three properties of $F$-statistics for inference about the mean vector in multiple regression and analysis of variance. The extra SSE due to imposing a set of linear conditions on the model tests the estimable part of those conditions. All other possible numerator SSs that test the same have not-lesser degrees of freedom and not-greater non-centrality parameter. 
 When factor-level combinations are coded by contrasts, the model restricted to eliminate an ANOVA effect is formulated by omitting that effect's columns from the model matrix. 
\end{abstract}

\section{Introduction}
Balanced-model two-factor Analysis of Variance (ANOVA) computations follow R. A. Fisher's prescription in his \emph{Statistical Methods for Research Workers} (see, e.g., Fisher (1938)). They are straightforward even for hand calculations and, as such, they are ubiquitous in textbooks and undergraduate and graduate courses on statistical methods. But unbalanced models are barely mentioned, and if they are, often it is to say that computations leading to sums of squares for main effects and interaction effects  are beyond the scope of the course.

The question, how (and even whether) to test main effects in models that permit interaction effects has been argued over almost since the beginnings of ANOVA nearly a century ago. Yates (1934) described the method of weighted squares of means (MWSM) to get a sum of squares to test main effects, and it has been the gold standard ever since. However, he did not describe what it tested, only that it ``provides an efficient estimate ... of the variance of the individual observations'' based on subclass averages, hence independent of error mean square.

Although there is practically unanimous agreement on the definition of main effects in balanced models, that has not been true of unbalanced models. Kutner (1974) listed three definitions. Speed, Hocking, and Hackney (1978) mentioned four. Francis (1973) noted differences in ANOVA sums of squares produced for the same data and model by four different statistical computing packages. Models built on dummy variables or with reference-level or effect coding of factor levels produced different sums of squares. 

SAS introduced Type III sums of squares in part to resolve this ambiguity, justified by asserting (Goodnight 1976) that ``[w]hen no missing cells exist ... Type III SS will coincide with Yate's weighted squares of means technique.'' Concluding that same paper, the author wrote: 

\begin{quotation}\noindent Perhaps (and just perhaps) we may someday be able to agree on the estimable functions we want to use in any given situation. If this day ever comes, we can then consolidate the different types of estimable functions (and live happily ever after).
\end{quotation}

Models for ANOVA effects fit readily into the framework of multiple regression models, of the form $\bm{\mu}=X\bm{\beta}$, where $\bm{\mu}$ is the mean vector of the $n$-variate response $\bm{Y}$. 
Once ANOVA effects are defined, as, say, $G\pr\bm{\beta}$, methods to formulate numerator sums of squares to test them have been known at least since the beginnings of ANOVA. One is as the extra SSE due to imposing the conditions $G\pr\bm{\beta} = \bm{0}$ on the model.  Fitting the restricted model can seem to be complicated and mysterious, particularly if it is done in terms of Lagrange multipliers and derivatives, as it is presented in most textbooks. It is simple and understandable, though, when it entails only omitting a set of predictor variables, e.g., $X_2\bm{\beta}_2 = \bm{0}$ in the model $X_1\bm{\beta}_1 + X_2\bm{\beta}_2$. 

Formulations of ANOVA effects for regression models typically are over-parameterized, and then inference is possible only about the estimable parts of them. While identifying those parts is tractable, doing so adds another step, algebraic or computational or pedagogical, to the process.

However a numerator SS is formulated, there is the question, whether another SS for the same effect might have better performance characteristics. In the conventional development of hypothesis tests, that becomes the question whether there is another SS that yields greater power for the same size. The salient properties of p-values, on the other hand, are captured in the CDFs of their probability distributions. Then one kind of p-value dominates another if its CDF is everywhere at least as great as the other's.

Three results are established here. Two pertain to the general linear model $X\bm{\beta}$. The first establishes that, for the proposition H$_0: G\pr\bm{\beta} = \bm{0}$, the Restricted Model - Full Model difference in $SSE$ (the RMFM SS) tests the estimable part, and only the estimable part, of $G\pr \bm{\beta}$. The second establishes that no other $SS$  is better than the RMFM SS for testing the same. The third establishes that, in models that include ANOVA effects, the correct restricted model for testing an effect can be formulated by omitting columns of $X$ corresponding to that effect.

The 2013 book by R. R. Hocking is arguably the most complete and detailed account of inference on ANOVA effects that is available today. For unbalanced two-factor models with no empty cells, it examines several different formulations of numerator SSs.  It notes that the methods produce different SSs, and concludes thereby that they test different hypotheses when their target is factor main effects. It describes the \emph{marginal means method} and notes in examples that it tests the target effect. With all cells filled, the model so formed has full column rank, and the equivalence of deleting columns and forcing equality of marginal means is apparent. 
The developments here encompass Hocking's account and establish further that,  even with empty cells, if factor levels are coded by contrasts, the extra SSE due to removing predictors corresponding to an ANOVA effect tests the estimable part of that effect,  and no other numerator SS is better.

\section{The Setting} 
The setting is the general multiple linear regression model. The random $n$-variate response $\bm{Y}$ has mean vector $\bm{\mu} = \E(\bm{Y}) = X\bm{\beta}$: that is, $\bm{\mu}\in\sp(X)$. The realized value of $\bm{Y}$ is $\bm{y}$. $X$ is a fixed, known $n\times (k+1)$ matrix with at least one row, at least one column, and at least one non-zero entry. Assume further that the distribution of $\bm{Y}$ is multivariate normal with variance-covariance matrix $\sigma^2\id_n$. The $(k+1)$-vector $\bm{\beta}$ and the positive scalar $\sigma^2$ are unknown parameters.

The set of all real $n$-vectors is denoted $\Re^n$. Vectors are column vectors, in boldface.  $S^\perp$ denotes the orthogonal complement of a subset $S$ of $\Re^n$. For sets $S_1$ and $S_2$ in $\Re^n$, $S_1 + S_2 = \{\bm{s}_1 + \bm{s}_2: \bm{s}_1\in S_1 \text{ and } \bm{s}_2\in S_2\}$. 
For matrices $A$ and $B$: $A\pr$, $(A, B) \equiv {\rm concat}(A, B)$, $\sp(A)$, $\tr(A)$, $\P_A$, $AB$, and $A\otimes B$ denote transpose, column-wise concatenation of matrices with the same number of rows, the linear subspace spanned by the columns of $A$, the trace of a square matrix, the orthogonal projection matrix onto $\sp(A)$, matrix product, and Kronecker product. ``If and only if'' is abbreviated iff. The $m\times m$ identity matrix is $\id_m$, and $\1_m$ denotes an $m$-vector of ones. 

Sums of squares -- SSs -- are prominent in methods of inference for such models. A sum of squares based on a matrix $M$ is $SS_M \equiv SS_M(\bm{y}) = \bm{y}\pr \P_M\bm{y}$.  Its degrees of freedom are $\nu_M = \tr(\P_M)$. Note that if $P$ is a symmetric idempotent matrix then $SS_P(\bm{y}) = \bm{y}\pr P\bm{y}$.

\section{The Conventional $F$-statistic}
Let $G$ be a given $(k+1)\times g$ matrix. Consider the proposition H$_0: G\pr\bm{\beta} = \bm{0}$. 
The conventional test statistic for assessing H$_0$ in light of $\bm{y}$ takes the general form
\beqn\label{conventional F}
F_P(\bm{y}) = \frac{SS_P/\nu_P}{SS_Q/\nu_Q},
\eeqn
where $P$ and $Q$ are symmetric idempotent matrices, $PQ = 0$, $\sp(Q)\subset\sp(X)^\perp$, $\nu_P = \tr(P)$, and $\nu_Q = \tr(Q)$. The focus here is on $P$, which (with $\sigma^2$) determines the non-centrality parameter and the numerator degrees of freedom of the distribution of $F_P$.  

With $\bm{Y} \sim \N(X\bm{\beta}, \sigma^2\id)$, $F_P$ follows an $F$-distribution with $\nu_P$ and $\nu_Q$ degrees of freedom and non-centrality parameter (ncp) $\delta_P^2(X\bm{\beta}) = SS_P(X\bm{\beta}/\sigma) = \bm{\beta}\pr X\pr PX\bm{\beta}/\sigma^2$. The ncp is zero, and the distribution is central, iff $\bm{\beta}$ is such that $PX\bm{\beta} = \bm{0}$: that is, iff $\bm{\beta}\in\sp(X\pr P)^\perp$.

For fixed numerator and denominator degrees of freedom and $\sigma^2$, the distribution of $F_P$ changes with, and only with, the ncp, hence with $PX\bm{\beta}$. In that case we shall say that $F_P$ \emph{tests} $PX\bm{\beta}$. Note that this is a narrow, specific definition of the word. 
Often ``tests H$_0$'' is vague, sometimes no more than an indication of intent. 
Here it means that the distribution of $F_P$ changes with $PX\bm{\beta}$, and any change in its distribution can result only from change in $PX\bm{\beta}$.
More generally, given  $G$ and  $P$, we shall say that $SS_P$ tests  $G\pr\bm{\beta}$ iff $\delta_P^2 = 0$ iff $G\pr\bm{\beta} = \bm{0}$. That in turn is equivalent to $\sp(G) = \sp(X\pr P)$.

Developments here are all in terms of testing propositions of the form H$_0: G\pr\bm{\beta} = \bm{0}$. Non-zero right-hand sides, like $G\pr\bm{\beta} = \bm{c}_0$, are accommodated by replacing $\bm{y}$ by $\bm{y}-X\bm{b}_0$, where $\bm{b}_0$ satisfies $G\pr\bm{b}_0 = \bm{c}_0$.

In a 1970 book and a 1973 article, B. K. Ghosh established fundamental properties of $F$ distributions. In Ghosh's notation, for $0<\alpha<1$, $\bar{F}_{\nu_P, \nu_Q ; \alpha}$ denotes the upper $\alpha$ quantile of the central $F$ distribution with $\nu_P$ and $\nu_Q$ degrees of freedom. 
Of particular importance here are that right-tail probabilities, of the form $\text{Pr}[ F_P(\bm{Y}) > \bar{F}_{\nu_P, \nu_Q; \alpha}\;|\; \nu_P, \nu_Q, \delta_P^2]$, are: monotone increasing in $\delta_P^2$; and, for $\delta^2_P>0$, monotone decreasing in $\nu_P$ and increasing in $\nu_Q$. For fixed $\alpha$, as a function of $\delta_P^2$, this is the power function of the size-$\alpha$ test that rejects H$_0: PX\bm{\beta} = \bm{0}$ when $F_P(\bm{y}) > \bar{F}_{\nu_P, \nu_Q; \alpha}$.
As a function of $\alpha$ for fixed $\delta_P^2$, it is the probability that a p-value from $F_P(\bm{y})$
takes a value $\leq \alpha$. Thus the CDF of the p-value is, at each $\alpha$,  increasing  in $\delta_P^2$; and, at each $\delta_P^2>0$, increasing in $\nu_Q$ and decreasing in $\nu_P$. When $\delta_P^2 = 0$, the distribution is uniform.

Given $P$ and $Q$ and fixed $\sigma^2$, if $SS_P$ tests $G\pr\bm{\beta}$ then the probability of lesser p-values increases as $G\pr\bm{\beta}$ departs from $\bm{0}$, and any shift in the distribution of p-values can be due only to departures of $G\pr\bm{\beta}$ from $\bm{0}$. That is, the distribution of p-values responds to, and only to, $G\pr\bm{\beta}$.

\section{Propositions}
Given $X$ and $\bm{g}\in\Re^{k+1}$, the linear function $\bm{g}\pr\bm{\beta}$ (or the vector $\bm{g}$) is defined to be \emph{estimable} iff $\bm{g}\in\sp(X\pr)$. For a $(k+1)\times g$ matrix $G$, the \emph{estimable part} of $\sp(G)$ is $\{\bm{g}\in\sp(G): \bm{g}\in\sp(X\pr)\} = \sp(X\pr)\cap\sp(G)$.

Let $N$ be a matrix such that $\sp(XN) = \{X\bm{b}: \bm{b}\in\Re^{k+1}\text{ and } G\pr\bm{b}=\bm{0}\}$. This is the model for $X\bm{\beta}$ restricted by the condition $G\pr\bm{\beta} = \bm{0}$. 
$N$ such that $\sp(N)=\sp(G)^\perp$ will do, but so will any $N$ such that $\sp(N)\supset\sp(G)^\perp$ and $\sp(N)\subset[\sp(G)\cap\sp(X\pr)]^\perp$. 

Proposition \ref{RMFM SS tests estimable part} establishes that there exists exactly one $SS_H$ with $\sp(H)\subset\sp(X)$ that tests the estimable part of $G\pr\bm{\beta}$.

\begin{propn}  
Let $H$ be a matrix with $\sp(H)\subset\sp(X)$. Then  $\sp(X\pr H) = \sp(X\pr)\cap\sp(G)$ iff $\P_H=\P_X - \P_{XN}$. \label{RMFM SS tests estimable part}
\end{propn}

\bigskip \noindent{\bfseries Proof.} \\\fbox{$\sp(H)\subset\sp(X)$ and $\sp(X\pr H) = \sp(G)\cap\sp(X\pr)$ $\imp$ $\P_H = \P_X - \P_{XN}$:} 

\noindent Show that $\sp(H) \subset \sp(\P_X - \P_{XN})$:
\beqnas
\sp(X\pr H) &=& \sp(X\pr)\cap\sp(G) \subset \sp(N)^\perp\\
\imp N\pr X\pr H &=& (XN)\pr H = \bm{0} \imp P_{XN}H = 0 \\
\imp (\P_X - \P_{XN})H &=& \P_X H = H \text{, because } \sp(H)\subset\sp(X),\\
\imp \sp(H) &\subset& \sp(\P_X - \P_{XN}).
\eeqnas
Show that $\sp(\P_X-\P_{XN})\subset\sp(H)$:  $\bm{z}\in\sp(\P_X - \P_{XN})$ $\imp$ $(XN)\pr\bm{z} = \bm{0}$ $\imp$ $X\pr\bm{z} \in\sp(X\pr)\cap\sp(N)^\perp \subset \sp(X\pr)\cap\sp(G) = \sp(X\pr H)$ $\imp$ $\exists$ $\bm{x}$ such that $X\pr\bm{z} = X\pr H\bm{x}$; since both $\bm{z}$ and $H\bm{x}$ are in $\sp(X)$, it follows that $\bm{z} = H\bm{x} \in \sp(H)$. Therefore $\sp(\P_X-\P_{XN}) \subset\sp(H)$, and therefore $\sp(H) = \sp(\P_X-\P_{XN})$. Both $\P_H$ and $\P_X-\P_{XN}$ are orthogonal projection matrices onto the same linear subspace, and therefore $\P_H=\P_X-\P_{XN}$.

\bigskip\noindent\fbox{$\P_H=\P_X-\P_{XN}$ $\imp$ $\sp(X\pr H) = \sp(G)\cap\sp(X\pr)$:}

\noindent If $\bm{z}\in\sp(X\pr H)$ then $\exists$ $\bm{x}$ such that $\bm{z}=X\pr H\bm{x}$; then $N\pr\bm{z} = (XN)\pr H\bm{x} = (XN)\pr \P_H H\bm{x}=\bm{0}$ $\imp$ $\bm{z} \in \sp(N)^\perp \subset \sp(G)$, and hence $\bm{z}\in\sp(X\pr)\cap\sp(G)$. Therefore $\sp(X\pr H) \subset\sp(X\pr)\cap\sp(G)$.

If $\bm{z}\in\sp(G)\cap\sp(X\pr)$, which is contained in $\sp(N)^\perp$, then $N\pr\bm{z}=\bm{0}$; and
$\exists$ $\bm{h}\in\sp(X)$  such that $\bm{z}=X\pr\bm{h}$, hence $N\pr X\pr\bm{h} = \bm{0}$ $\imp$ $\bm{h}\in\sp(X)\cap\sp(XN)^\perp = \sp(\P_H)=\sp(H)$, and therefore $\bm{z}=X\pr\bm{h}$ is in $\sp(X\pr H)$. Therefore $\sp(G)\cap\sp(X\pr) \subset\sp(X\pr H)$, and therefore $\sp(X\pr H) = \sp(G)\cap\sp(X\pr)$. \hfill $\blacksquare$

\bigskip
The unique symmetric idempotent matrix $\P_H$ with $\sp(H)\subset\sp(X)$ such that   $\delta_P^2 = 0$ iff the estimable part of $G\pr\bm{\beta}$ is $\bm{0}$ is  $\P_H = \P_X - \P_{XN}$. Then
\beqna
SS_H(\bm{y}) &=& \bm{y}\pr(\P_X - \P_{XN})\bm{y}\nnlf
& =& \bm{y}\pr(\id-\P_{XN})\bm{y} - \bm{y}\pr(\id-\P_X)\bm{y}\nnlf
& =& SSE_{XN} - SSE_{X},
\eeqna
which is the increase in $SSE$ due to restricting the model $X\bm{\beta}$ by $G\pr\bm{\beta} = \bm{0}$. Call this the Restricted Model -- Full Model (RMFM) SS for $G\pr\bm{\beta}$. All that is required to get this SS is to get the right restricted model, which requires $N$. 

It is not necessary to identify the estimable part of $G$ in order to get $SS_H$. However, in order to properly interpret the results, one needs to know what $SS_H$ tests. One way is to find the estimable part directly from $\P_H$ by finding $G_H$ such that $\sp(G_H) = \sp(X\pr \P_H)$.  This can entail searching for fewer recognizable linear combinations of the columns of $X\pr H$ that span the same space.

In the construction of $N$, and hence of $\P_H$, it is important to keep in mind that all that matters about $G$ is its column space. The set $\{\bm{\beta}\in\Re^{k+1}: G\pr\bm{\beta} = \bm{0}\}$ is the same for any matrix $G_*$ such that $\sp(G_*) = \sp(G)$. $N$ also can come in many forms; all that is required is that $\sp(G)^\perp\subset\sp(N) \subset \sp(G)^\perp + \sp(X\pr)^\perp$. 

For a matrix $C$, let $\nu_C = \dim\sp(C) = \tr(\P_C)$. Proposition \ref{Q and H} resembles corresponding properties of solution sets of linear equations $A\bm{x} = \bm{b}$: the unique solution $\bm{x}_0$ in $\sp(A\pr)$ has minimum norm; the solution set can be expressed as $\bm{x}_0 + \sp(A\pr)^\perp$; and all solutions have the same orthogonal projection $\bm{x}_0$ in $\sp(A\pr)$. It is to non-centrality parameters in this setting as the Gauss-Markov Theorem is to variances of unbiased linear estimators.

\begin{propn}  \label{Q and H} Let $X$, $H$, and $L$ be matrices with $n$ rows such that $\sp(H)\subset\sp(X)$ and  $\sp(X\pr L) = \sp(X\pr H)$. Then  
\begin{enumerate}
\item $\sp(\P_XL) = \sp(H)$,
\item $\sp(L)\subset\sp(H) + \sp(X)^\perp$, 
\item $X\pr \P_HX - X\pr \P_L X$ is nnd, and
\item $\nu_H = \nu_{\P_X L} \leq \nu_L \leq \nu_H + n-\nu_X$.
\end{enumerate}
\end{propn}

{\bfseries Proof:} \begin{enumerate}
\item \fbox{$\sp(\P_XL) \subset\sp(H)$:} $\bm{z} = \P_XL\bm{a} \in \sp(\P_XL)$ $\imp$ $\exists$ $\bm{b}$ such that $X\pr \bm{z} = X\pr H\bm{b}$ $\imp$ $\bm{z}-H\bm{b} \in \sp(X)\cap\sp(X)^\perp$ $\imp$ $\bm{z}=H\bm{b} \in \sp(H)$.

\fbox{$\sp(H)\subset\sp(\P_XL)$:} $\bm{z}=H\bm{b} \in \sp(H)$ (hence $\bm{z}\in\sp(X)$) $\imp$ $\exists$ $\bm{a}$ such that $X\pr\bm{z} = X\pr H\bm{b} = X\pr L\bm{a} = X\pr \P_XL\bm{a}$ $\imp$ $\bm{z}-\P_XL\bm{a}\in \sp(X)\cap\sp(X)^\perp$ $\imp$ $\bm{z}\in \sp(\P_X L)$. Therefore $\sp(\P_XL) = \sp(H)$.

\item If $\sp(X\pr L) = \sp(X\pr H)$, then $L\bm{a}\in \sp(L)$ $\imp$ $\exists$ $\bm{b}$ such that $X\pr L\bm{a} = X\pr H\bm{b}$ $\imp$ $L\bm{a} - H\bm{b} \in \sp(X)^\perp$ $\imp$ $L\bm{a} \in \sp(H) + \sp(X)^\perp$. 

\item Because $\sp(H)$ and $\sp(X)^\perp$ are orthogonal, the orthogonal projection matrix onto $\sp(H)+\sp(X)^\perp$ is $\P_H + (\id-\P_X)$. And, because $\sp(L)\subset\sp(H) + \sp(X)^\perp$, $Q = \P_H+(\id-\P_X) - \P_L$ is symmetric and idempotent. It follows then that, for any vector $\bm{z}$, 
\beqna 
\bm{z}\pr (X\pr \P_H X - X\pr \P_LX)\bm{z} &=& \bm{z}\pr X\pr QX\bm{z} \nnlf
&=&
(Q\pr X\bm{z})\pr (Q\pr X\bm{z}) \geq 0,
\eeqna 
that is, that $X\pr \P_H X - X\pr\P_L X$ is nnd. 

\item Because $Q = QQ\pr$, $\tr(Q) = \tr(QQ\pr) \geq 0$, and hence 
$\nu_L =\tr(\P_L) \leq \tr(\P_H) + \tr(\id-\P_X) = \nu_H + n-\nu_X$. Because $\sp(H) = \sp(\P_XL)$, $\nu_H = \nu_{\P_XL} \leq \nu_L$.
\hfill$\blacksquare$
\end{enumerate}

\bigskip While $SS_H$ and $SS_L$ both test $H\pr X\bm{\beta}$, $\delta_L^2 \leq \delta_H^2$ and $\nu_L \geq \nu_H$. Assume for now that $F_H$ and $F_L$ both use the same denominator SS and that it is independent of both $SS_H$ and $SS_L$. (This could leave it with fewer denominator degrees of freedom than $\tr(\id-\P_X)$.) Recalling Ghosh's results (1970, 1973), this means that, when $H\pr X\bm{\beta} \neq \bm{0}$, the CDF of p-values is everywhere $\geq$ for $F_H$ than for $F_L$. In this sense, $F_H$ is \emph{as good as} $F_L$ for any $F_L$ that tests the estimable part of $G\pr\bm{\beta}$.

Let $X_0$ be a matrix such that $\sp(X_0)\subset\sp(X)$. Let $H$ be a matrix such that $\sp(H)\subset\sp(X)$. Then $SS_H$ is the best numerator SS for $H\pr\bm{\mu}$ in the model $\sp(X)$.  Let $H_0 = \P_{X_0}H$. In the sub-model $\sp(X_0)$, $SS_H$ and $SS_{H_0}$ test the same hypothesis, because $\sp(X_0\pr H) = \sp(X_0\pr H_0)$, but $SS_{H_0}$ is better than $SS_H$ if $\sp(H_0) \neq \sp(H)$.

\section{Models for ANOVA Effects}\label{ANOVA models}
The third proposition has to do with models for factor effects in a two-factor setting. The factors are named A and B. Factor A has $a$ levels and factor B has $b$ levels. There are $ab$ factor-level combinations (FLCs). Let $\bm{\eta}$ denote the $ab$-vector of \emph{cell means}: that is, $\bm{\eta} = (\eta_{ij}: i=1, \ldots, a, j=1, \ldots, b)$, where $\eta_{ij}$ is the expected value of the response $Y$ under the $i, j$ FLC.  The following definitions and notation are intended to extend readily to settings with more than $f=2$ factors.

For positive integers $m$ define $U_m = (1/m)\1_m\1_m\pr$ and $S_m = \id - U_m$,  the orthogonal projection matrices onto $\sp(\1_m)$ and $\sp(\1_m)^\perp$, respectively.  Let $\bc{B}^2=\{00, 10, 01, 11\}$ denote the set of all binary pairs, $\bm{j}=j_1j_2$ with each $j_k\in\{0,1\}$. Let $a_1 = a$, $a_2 = b$, and define matrices $H_{\bm{j}}$ by:
\beqn
H_{\bm{j}} = \bigotimes_{k=1}^2\left\{\begin{array}{l} U_{a_k} \text{ if }j_k=0,\\
S_{a_k} \text{ if } j_k = 1. \end{array}\right.
\eeqn
Then H$_{00} = U_a\otimes U_b$, $H_{10} = S_a\otimes U_b$, $H_{01} = U_a\otimes S_b$, and $H_{11} = S_a\otimes S_b$. Note that these four matrices are symmetric, idempotent, pairwise orthogonal, and their sum is $\id_a\otimes \id_b$. A sum of any subset of these matrices is symmetric and idempotent, and it is orthogonal to any $H_{\bm{j}}$ not in the sum. 

Define factor A \emph{main effects} to be linear functions of $(\bar{\eta}_{i\cdot} - \bar{\eta}_{\cdot\cdot}) = H_{10}\bm{\eta}$. There are no A main effects iff $H_{10}\bm{\eta}=\bm{0}$. Define B main effects and AB interaction effects similarly by  $H_{01}\bm{\eta} = (\bar{\eta}_{\cdot j}-\bar{\eta}_{\cdot\cdot})$ and $H_{11}\bm{\eta} = (\eta_{ij} - \bar{\eta}_{i\cdot} - \bar{\eta}_{\cdot j} + \bar{\eta}_{\cdot\cdot})$, respectively.

 Factor-effects models for $\bm{\eta}$ are linear subspaces spanned by subsets of these four matrices. The model that includes all effects is $\Re^{ab} = \sp(H_{00}+H_{10}+H_{01}+H_{11})$; that excludes AB interaction effects, $\sp(H_{00}+H_{10}+H_{01})$; that excludes A main effects, $\sp(H_{00}+H_{01}+H_{11})$; and so on.  Note further that, for example, $\sp(H_{00}+H_{10}+H_{01}) = \sp(H_{00},H_{10},H_{01})$, the latter formed by concatenating the three matrices column-wise; and that the result can also be expressed as $\sp(H_{00}) + \sp(H_{10}) + \sp(H_{01})$.

For positive integers $m$, let $C_m$ denote a matrix such that $\sp(C_m) = \sp(\1_m)^\perp$.  Columns of $C_m$ are contrasts that span $\{\bm{c}\in\Re^m: \1_m\pr\bm{c}=0\}$. 
Then also $\sp(C_m) = \sp(S_m)$ and $\P_{C_m} = S_m$. One possible choice for $C_m$ is $S_m$. In what follows, each appearance of $C_m$ may have a different choice of columns.  
For each binary pair $\bm{j}\in\bc{B}^2$, define
\beqn C_{\bm{j}} = \bigotimes_{k=1}^2\left\{ \begin{array}{l} \1_{a_k} \text{ if } j_k=0,\\ C_{a_k} \text{ if } j_k=1. \end{array}
   \right.
\eeqn

Recall that, for matrices $A$ and $B$, $\P_{A\otimes B} = \P_A\otimes \P_B$, linear subspaces and their orthogonal projection matrices are one-to-one, and $\sp(\P_A) = \sp(A)$. It follows that $\P_{C_{\bm{j}}} = H_{\bm{j}}$ and hence 
$\sp(C_{\bm{j}}) = \sp(H_{\bm{j}})$. Further, for example, $\sp(H_{00}+H_{01}+H_{11}) = \sp(H_{00},H_{01},H_{11}) = \sp(C_{00}) + \sp(C_{01}) + \sp(C_{11}) = \sp(C_{00}, C_{01}, C_{11})$ is a model that excludes only A main effects. Omitting the columns  corresponding to one or more effects constrains the model for the cell means to exclude those and only those effects. Proposition \ref{effects models} follows from the properties already noted. Extending it to  $f$  factors is straightforward upon replacing $\bc{B}^2$ by $\bc{B}^f$.

\begin{propn}\label{effects models}
Let $\bc{J}$ be a non-empty subset of $\bc{B}^2$. Let $H_{\bc{J}} = \sum\{H_{\bm{j}}: \bm{j}\in\bc{J}\}$. Let $C_{\bc{J}} = {\rm concat}\{C_{\bm{j}}: \bm{j}\in\bc{J}\}$. Then $\sp(C_{\bc{J}}) = \sp(H_{\bc{J}})$. And, for $\bm{j}_*\in\bc{J}$, $\sp(C_{\bc{J}})\cap\sp(H_{\bm{j}_*})^\perp = \sp(C_{\bc{J}\backslash\bm{j}_*})$.
\end{propn}

\bigskip
In the customary dot and bar notation for subscripted terms, a dot replacing a subscript signifies summation, and a bar signifies the average, over the range of the subscript. Thus, as examples, $n_{i\cdot} = \sum_{j}n_{ij}$ and $\bar{\eta}_{i\cdot} = (1/b)\sum_{j}\eta_{ij}$. 

Consider now that $n_{ij}$ responses $y_{ijs}$ are observed under each factor-level combination $i,j$. Assume that $n_{i\cdot} > 0$ and $n_{\cdot j} >0$, but that cells with $n_{ij}=0$ are not excluded. Let $n \equiv n_{\cdot\cdot}$ and define the $n\times  ab$ matrix $\mbk$ to have, in its $i,j,s$-th row, 1 in the $i,j$-th column and $0$s in all other columns. Then the $i,j$-th column of $\mbk$ has exactly $n_{ij}$ 1s, and each row of $\mbk$ has exactly one 1. Note that if $n_{ij}=0$ then the $i,j$-th column of $\mbk$ is $\bm{0}_n$.  

Let $\bc{J}$ be a non-empty subset of $\bc{B}^2$. With the effects in $\bc{J}$, one way to express the model for the cell means is $\bm{\eta}\in\sp(C_{\bc{J}})$. The model for the mean vector  $\bm{\mu}=(\mu_{ijs}=\E(Y_{ijs}))$ can be expressed as $\bm{\mu}\in\sp(\mbk C_{\bc{J}})$.

For example, consider the saturated model, with $\bc{J}=\bc{B}^2$. The model for the cell means is $\bm{\eta}\in\Re^{ab}$. In terms of $C_{\bc{J}}$ it is $\sp(C_{\bc{J}})$. The model for $\bm{\mu} = \mbk\bm{\eta}$ is $\sp(\mbk C_{\bc{J}})$, which in this case is equivalent to $\sp(\mbk)$. Let $\bm{j}_* = 10$. For the proposition $H_{10}\bm{\eta} = \bm{0}$, that there are no A main effects, by Proposition \ref{effects models}, the restricted model for $\bm{\eta}$ is $\sp(C_{\bc{J}\backslash\bm{j}_*}) = \sp(C_{00}, C_{01}, C_{11})$. Then the restricted model for $\bm{\mu}$ is $\sp(\mbk C_{\bc{J}\backslash \bm{j}_*})$, deleting the columns corresponding to A main effects, $\mbk C_{10}$, from the full model. The RMFM SS for A main effects is then
\beqn
SS_{10} = \bm{y}\pr (\P_{\mbk C_{\bc{J}}} - \P_{\mbk C_{\bc{J}\backslash \bm{j}_*}})\bm{y}.
\eeqn
It is the difference in SSE for the regression of $\bm{y}$ on the restricted model, $\mbk(C_{00},C_{01},C_{11})$, and the full model, $\mbk$ or $\mbk(C_{00}, C_{10}, C_{01}, C_{11})$. By Proposition \ref{RMFM SS tests estimable part}, its ncp is $0$  iff all estimable A main effects are $\bm{0}$.

Everything is simpler in balanced models with $m$ observations per cell. There $\mbk = \id_{ab}\otimes \1_m$ (possibly after re-arranging rows).
Then for any model $\bc{J}$, $\P_{\mbk C_{\bc{J}}} = H_{\bc{J}}\otimes U_m$; and for any effect $\bm{j}_*$ in $\bc{J}$, the matrix of the RMFM SS for $\bm{j}_*$ is 
$\P_{\mbk H_{\bm{j}_*}} = H_{\bm{j}_*}\otimes U_m$, which tests $(H_{\bm{j}_*}\otimes U_m)(H_{\bc{J}}\otimes U_m)(\bm{\eta}\otimes \1_m) = (H_{\bm{j}_*}\bm{\eta})\otimes \1_m = \bm{0}$.
For example, the RMFM SS for testing A main effects is $\bm{y}\pr (H_{10}\otimes U_m)\bm{y} = m\sum_{i=1}^a\sum_{j=1}^b(\bar{y}_{i\cdot\cdot} - \bar{y}_{\cdot\cdot\cdot})^2$, the same as the A SS as if there were no factor B.


\section{Examples} The purpose of these examples is to illustrate the consequences of the three propositions. Computational results are given so that the interested reader can verify them. Basic computational tools for orthogonal projection and other operations on linear subspaces follow from the Gram-Schmidt construction as described in LaMotte (2014).

Consider three configurations $N = (n_{ij})$ of subclass numbers for models with $a=b=3$. The $n_{ij}$s are arranged in $3\times 3$ arrays corresponding to the same arrangement of cell means:
\beqns
N_0 = \left(\begin{array}{ccc}1 & 2 & 3\\3 & 1 & 2\\3 & 2 & 1\end{array}\right),\;
N_1 = \left(\begin{array}{ccc}6 & 4 & 4\\3 & 2 & 2\\3 & 2 & 2\end{array}\right),\;
N_2 = \left(\begin{array}{ccc}0 & 2 & 3\\3 & 1 & 2\\3 & 2 & 1\end{array}\right).\;
\eeqns
$N_0$ is unbalanced with no special features, $N_1$ has a property called \emph{proportional subclass numbers} (psn), and $N_2$ has an empty cell.

For each configuration, consider three models, $ m = 1, 2, 3$, for $\bm{\mu}=\mbk\bm{\eta}$, all including A main effects: 
\begin{enumerate}
  \item $\sp[\mbk(C_{00},C_{10})]$, including only an intercept and A main effects,
  \item $\sp[\mbk(C_{00}, C_{10}, C_{01})]$, the additive model, excluding only interaction effects, and 
  \item $\sp[\mbk(C_{00}, C_{10}, C_{01}, C_{11})] = \sp(\mbk)$, the saturated model.
\end{enumerate}
With $a = b = 3$, $C_3 = \left(\begin{array}{rr}2 & 0\\-1 & 1\\-1 & -1\end{array}\right)$ was used to define $C_{\bm{j}}$, $\bm{j} = j_1j_2 \in\bc{B}^2$ for the results shown here. In particular, $C_{10} = C_3\otimes \1_3$, a $9\times 2$ matrix in which each row of $C_3$ is repeated $b=3$ times.

Each configuration of subclass numbers specifies a different $\mbk$. The focus of these examples is on testing A main effects, $H_{10}\bm{\eta}$ or $C_{10}\pr\bm{\eta}$. 
    
Each model has the form $\sp(\mbk M)$, and $\sp(M)$ contains all A main effects contrasts.  $SS_P$ tests $P\mbk M\bm{\tau}$. In terms of the cell-means vector $\bm{\eta} = M\bm{\tau}\in\sp(M)$, it tests $G\pr\bm{\eta}$, where $\sp(G) = \sp(\P_M\mbk\pr P)$. Here and in what follows,  $G$ is not unique and may be replaced by any matrix with the same column space.

The restricted model that excludes all estimable A main effects is formulated in each case by removing the $\mbk C_{10}$ columns, by Proposition \ref{effects models}.
As $\P_M\mbk\pr P$ is $ab\times n$, $G$  can be chosen with fewer columns (down  to $\tr(P)$) in order to express the conditions in terms of recognizable contrasts. 

Denote the RMFM SSs for A main effects in models 1-3 by $SS_{tA}$, $t = 1, 2, 3$ ($t$ for \emph{type} of SS), respectively. Denote the matrices of these quadratic forms correspondingly by $P_{tA}$. In SAS's (1978) widely-used nomenclature, $SS_{1A}$ and $SS_{2A}$ are Type I and Type II SSs, respectively. It can be shown that, under subclass numbers $N_0$ and $N_1$,  $SS_{3A}$ is the same as Yates's (1934) MWSM sum of squares.

If the subclass numbers were all equal then the three SSs would be the same, and they would test A main effects in all three models.   

The question to be investigated here is what each $SS_{tA}$ tests in each model. This can be accomplished by finding $G_{tm}$ such that $\sp(G_{tm}) = \sp(\P_{M_m}\mbk\pr P_{tA})$ for each pair $t, m$, so that $SS_{tA}$ tests $G_{tm}\pr\bm{\eta}$ in model $m$. By Proposition \ref{RMFM SS tests estimable part}, $SS_{mA}$ tests the estimable part of $H_{10}\bm{\eta}$ in model $m$, and so $\sp(G_{mm})\subset\sp(H_{10})$. Then $G_{21} = \P_{M_1}G_{22} = G_{22}$, because $\sp(M_1)\supset\sp(H_{10})$; and similarly $G_{31} = G_{32} = G_{33}$. Once $G_{11}$, $G_{22}$, and $G_{33}$ are determined, then, only $G_{12}$, $G_{13}$, and $G_{23}$ remain to be found.

Verify computationally that, under both $N_0$ and $N_1$, $\sp(\P_{M_m}\mbk\pr P_{mA}) = \sp(C_{10})$ in all three models, and hence $G_{11} = G_{22} = G_{33} = C_{10}$. Under $N_2$, verify that the same  holds in models 1 and 2, and $G_{11} = G_{22} = C_{10}$, but that $\sp(\P_{M_3}\mbk\pr P_{3A})$ is spanned by the second column of $C_{10}$ alone, so $G_{33} = (0 , 1, -1)\pr \otimes \1_3$, and $G_{31} = G_{32} = G_{33}$. For model 3 under $N_2$, $SS_{3A}$ tests $G_{33}\pr\bm{\eta}$, which is proportional to $\bar{\eta}_{2\cdot} - \bar{\eta}_{3\cdot}$; and it tests the same under models 1 and 2.  For comparison, Hocking (2013, p. 324), citing Hocking, Hackney, and Speed (1978), gives formulas for $G_{23}$ and $G_{13}$ in terms of the $n_{ij}$s.

Table \ref{tab 1} shows $G_{12}$, $G_{13}$, and $G_{23}$ under the three sets of subclass numbers. In balanced settings these would all be $C_{10}$ or its equivalent. It seems to be widely thought  to be known that the same holds in psn settings (like $N_1$), that is, that $SS_{1A}$, which tests A main effects in model 1, also tests A main effects in models 2 and 3. In Table \ref{tab 1}, under $N_1$, $G_{12} = C_{10}$, and so $SS_{1A}$ does test $H_{10}\bm{\eta}$ in model 2; but it does not in model 3, and in fact $\sp(G_{13})\cap\sp(H_{10}) = \{\bm{0}\}$: in model 3, $SS_{1A}$ does not test any part of A main effects, despite the fact that $N_1$ has proportional subclass numbers.

\begin{table}
\begin{center}
\begin{tabular}{|rrr|rrr|rrr|}\hline
\multicolumn{3}{|c}{$G_{12}\pr$}&\multicolumn{3}{|c}{$G_{13}\pr$} & \multicolumn{3}{|c|}{$G_{23}\pr$}\\\hline
\multicolumn{9}{|c|}{$N_{0}$}\\\hline
2 & 3 & 4 & 1 & 2 & 3 & 27 & 42 & 42\\
-1 & 0 & 1 & 0 & 0 & 0 & 18 & 0 & -14\\
-4 & -3 & -2 & -3 & -2 & -1 & -45 & -42 & -28\\\hline
-2 & -4 & -3 & -1 & -2 & -3 & -255 & -246 & -585\\
7 & 5 & 6 & 6 & 2 & 4 & 960 & 452 & 760\\
-2 & -4 & -3 & -3 & -2 & -1 & -705 & -206 & -175\\\hline
\multicolumn{9}{|c|}{$N_{1}$}\\\hline
2 & 2 & 2 & 18 & 12 & 12 & 18 & 12 & 12\\
-1 & -1 & -1 & -3 & -2 & -2 & -3 & -2 & -2\\
-1 & -1 & -1 & -15 & -10 & -10 & -15 & -10 & -10\\\hline
0 & 0 & 0 & -3 & -2 & -2 & -3 & -2 & -2\\
1 & 1 & 1 & 6 & 4 & 4 & 6 & 4 & 4\\
-1 & -1 & -1 & -3 & -2 & -2 & -3 & -2 & -2\\\hline
\multicolumn{9}{|c|}{$N_{2}$}\\\hline
45 & 95 & 130 & 0 & 36 & 54 & 0 & 12 & 12\\
-39 & 11 & 46 & 3 & 1 & 2 & 9 & 0 & -4 \\
-141 & -91 & -56 & -48 & -32 & -16 & -9 & -12 & -8 \\\hline
-5 & -14 & -11 & 0 & -12 & -18 & 0 & -156 & -315\\
25 & 16 & 19 & 30 & 10 & 20 & 360 & 212 & 370\\
-5 & -14 & -11 & -15 & -10 & -5 &  -360 & -56 & -55\\\hline
\end{tabular}
\end{center}
\caption{$G_{12}$, $G_{13}$, and $G_{23}$ for subclass numbers $N_0$, $N_1$, and $N_2$ such that $\sp(G_{tm}) = \sp(\P_{M_m}\mbk\pr P_{tA})$. The two rows of $G_{tm}\pr$ (columns of $G_{tm}$) are shown as $a\times b = 3\times 3$ arrays corresponding to cells $i,j$.}  \label{tab 1}
\end{table}

\section{Comments and Conclusions}
Practically all of the formulations shown here for ANOVA models have been around in one form or another for a long time, including the designation of  ANOVA effects by binary tuples $\bm{j}$ and sets of effects by subsets of those tuples, the basic projection matrices $S_m$ and $U_m$, and projection matrices $H_{\bm{j}}$ to express effects in terms of the cell means. Coding factor levels in terms of contrasts has been used widely to construct $X$ with full column rank and to avoid the ``containment'' properties of dummy-variables coding. Francis (1973, Section 4), for example, shows that the computing package BMDX64 used what is often called ``effect'' coding. The extra SSE for numerator SS has been standard practice, although the form $(G\pr\hat{\bm{\beta}})\pr [\var(G\pr\hat{\bm{\beta}})/\sigma^2]^-(G\pr\hat{\bm{\beta}})$ seems to be preferred. They are the same if $G\pr\bm{\beta}$ is estimable, but the second form tests more than the estimable part of $G\pr\bm{\beta}$ otherwise. 

The methods, models, and computational processes for multiple regression and ANOVA models are long-established and well-known. As far as I have been able to tell, if the three results established here are known, they are not widely recognized. In some cases, they provide justification for methods that have long been accepted and taught as standard procedure, but whose properties have not been rigorously established. And they might lead to recognition that some methods that were developed specifically to handle the complications due to unbalancedness and empty cells in models for ANOVA effects -- like Yates's MWSM and SAS's Type III and Type IV functions -- were unnecessary.

It can be shown that Proposition \ref{effects models} holds as well in models that include covariates and covariate-by-factor interaction effects. Generally, then, if factor-level combinations are coded in terms of contrasts, any effect in the model can be tested with an RMFM numerator SS where the correct restricted model results from omitting the corresponding set of columns from the $X$ matrix.  The resulting $F$-statistic tests the estimable part of the effect. And, within the assumed model, no other numerator SS that tests the same functions yields a more favorable ncp or numerator degrees of freedom.

\section{Bibliography}
\begin{itemize}\setlength{\topsep}{0cm}\setlength{\labelsep}{0cm}
\setlength{\itemsep}{0cm}\setlength{\parsep}{0cm}
\setlength{\parskip}{0cm}\setlength{\itemindent}{-1cm}
\item[] Fisher, R. A. (1938).  Statistical Methods for Research Workers, 7th Edition.  Oliver and Boyd, London.
\item[] Francis, I. (1973). A comparison of several analysis of variance programs. Journal of the American Statistical Association 68: 860-865.
\item[] Ghosh, B. K. (1970). Sequential Tests of Statistical Hypotheses. Reading, Mass.: Addison-Wesley.
\item[] Ghosh, B. K. (1973). Some Monotonicity Theorems for $\chi^2$, $F$, and $t$ Distributions with Applications. Journal of the Royal Statistical Society, Series B (Methodological), 35(3): 480-492.
\item[] Goodnight, J. H. (1976). The General Linear Models procedure. Proceedings of the First International SAS User's Group.  SAS Institute Inc., Cary, NC.
\item[] Hocking, R. R. (2013). Methods and Applications of Linear Models, Third Edition.  John Wiley \& Sons, Inc., Hoboken, New Jersey.
\item[] Hocking, R. R., Hackney, O. P., and Speed, F. M. (1978). The analysis of linear models with unbalanced data. In: David, H. A., editor, Contributions to Survey Sampling and Applied Statistics: Papers in Honor of H. O. Hartley. New York: Academic Press.
\item[] Kutner, M. H. (1974). Hypothesis testing in linear models (Eisenhart Model I). The American Statistician, 28(3): 98-100.
\item[] LaMotte, L. R. (2014). The Gram-Schmidt construction as a basis for linear models. The American Statistician 68(1): 52-55.
\item[] SAS Institute Inc. (1978).  SAS Technical Report R-101, Tests of hypotheses in fixed-effects linear models.  Cary, NC.
\item[] Speed, F. M., Hocking, R. R., and Hackney, O. P. (1978). Methods of analysis of linear models with unbalanced data. Journal of the American Statistical Association 73: 105-112.
\item[] Yates, F. (1934).  The analysis of multiple classifications with unequal numbers in the different classes.  Journal of the American Statistical Association, 29(185): 51-66.

\end{itemize}

\end{document}